\documentclass{elsart5p}

\def\RR{{\rm I\kern-.1567em R}}                              % Doppel R
\def\ZZ{{\sf Z\kern-4.5pt Z}}

\usepackage{graphics, graphicx}
\begin{document}
\begin{frontmatter}

\title{Compact shell solitons in $K$ field theories}

%\vspace*{1cm}

\author[san]{C. ~Adam}
\ead{adam@fpaxp1.usc.es}
\author[san]{P. ~Klimas }
\ead{klimas@fpaxp1.usc.es}
\author[san]{J. ~S\'{a}nchez-Guill\'{e}n}
\ead{joaquin@fpaxp1.usc.es}

\address[san]{Departamento de Fisica de Particulas, Universidad
       de Santiago and Instituto Galego de Fisica de Altas Enerxias
       (IGFAE) E-15782 Santiago de Compostela, Spain}

\author[were1,were2]{A. ~Wereszczy\'{n}ski}
\ead{wereszczynski@th.if.uj.edu.pl}

\address[were1]{The Niels Bohr Institute, Copenhagen University, 
Blegdamsvej 17, DK-2100 Copenhagen {\O},
       Denmark}

\address[were2]{Institute of Physics,  Jagiellonian University,
       Reymonta 4, 30-059 Krak\'{o}w, Poland
}

\begin{abstract}
Some models providing shell-shaped  
static solutions with compact support (compactons) 
in 3+1 and 4+1 dimensions are introduced, and the corresponding exact
solutions are calculated analytically. These solutions turn out to be 
topological solitons, and may be classified as maps $S^3 \to S^3$ and
suspended Hopf maps, respectively.  
The Lagrangian of these models is given by a scalar field with a non-standard
kinetic term ($K$ field) coupled to a pure Skyrme 
term restricted to $S^2$, rised to the appropriate power to avoid the Derrick
scaling argument. 
Further, the existence of infinitely many exact shell solitons is explained using the generalized integrability approach. Finally,
similar models allowing for non-topological 
compactons of the ball type in
3+1 dimensions are briefly discussed.
\end{abstract}


\end{frontmatter}
%\twocolumn 

\section{Introduction}
Recently, some effort has been invested into the investigation of 
compactons, that is, soliton solutions of non-linear field theories with
compact support. By now, two established classes of scalar field theories 
are known which give rise to the existence of compacton solutions. One may
either choose potentials in the Lagrangian (or energy density) which have a
non-continuous first derivative at (some of) their minima
\cite{arodz1} - \cite{kuru}, or one may employ
a non-standard kinetic term ($K$ field theory) \cite{comp}, \cite{bazeia1}.
Concretely, the kinetic term has to contain higher than second powers in the
first derivatives of the fields. 
For nonrelativistic field theories, compactons were first discovered and
studied for some generalizations of the KdV equation in 
\cite{RoHy}, \cite{CSS}.
Compactons can be realized directly in some
mechanical systems \cite{arodz1}, and, in addition, they have been applied 
recently to brane cosmology \cite{comp-br1}, \cite{comp-br2}, \cite{bazeia2}. 
Most of these investigations have dealt with topological
compactons, where the existence and stability of the compacton solutions is
related to a nontrivial vacuum manifold and some nonzero topological charge.
Also in this letter we shall 
mainly deal with this case of topological compactons, which turn out to have
the shape of shells, and only at the end we will investigate a similar theory
which, however, gives rise to non-topological compactons with a ball shape. 

As is typical in soliton theory in general, it is easier to find systems with
compacton solutions in low (that is 1+1) dimensions. The simplest, most
obvious generalization of topological compacton systems 
with a non-standard kinetic term to higher dimensions 
meets the
same obstacles as in the case of conventional solitons, and also the remedy to
circumvent the obstacle is the same, namely the introduction of a gauge field
in addition to the scalar fields,
like in the case of vortices and monopoles \cite{comp-vort}.     
In this letter, we shall pursue a different path for the construction of
higher-dimensional topological compactons. Concretely, we couple a real scalar
field $\xi$ with non-standard kinetic term to a complex scalar field $u$
where $u$ maps from a compact submanifold of the base space ($S^2$ or $S^3$
for $\RR^3$ or $\RR^4$, respectively) to an $S^2$ target space, thereby
providing an important contribution to the non-trivial topology of the
compacton solutions. The non-trivial topology is completed by the boundary
conditions on $\xi$ required by finiteness of the energy.
For the model
constructed in Section 2, the 
 complex scalar field $u$ has the topology of a Hopf map $S^3 \to S^2$, 
and $\xi$ provides the suspension of this Hopf map to a map $S^4 \to S^3$. 
The base space for this example is, therefore, 4+1 dimensional.
In Section 3, we discuss similar models which allow for
compactons in 3+1 dimensions, where $u$ describes a map $S^2 \to S^2$.
In Section 3.1, we study a model which again gives rise to analytical
solutions which are shell-shaped.
In Section 3.2 we introduce a slightly different model, where it turns out
that the compact solitons are ball-shaped and, further, are no longer
topological.

\section{4-dimensional compactons}  

The specific $4+1$ dimensional model we are going to consider is given by 
the following expression
\begin{equation}
L= |\xi_{\nu} \xi^{\nu}| \xi_{\mu} \xi^{\mu} - \sigma (\xi) H_{\mu \nu}^2 ,  
\label{model}
\end{equation}
where $\xi_\mu \equiv \partial_\mu \xi$ etc.
It describes a real scalar field $\xi$ with a non-standard kinetic term
coupled to the pure Skyrme 
term constrained to $S^2$ target space
\begin{equation}
H_{\mu \nu}^2 = \frac{1}{(1+|u|^2)^4} \left[ (u_{\mu}\bar{u}^{\mu})^2 
- u_{\mu}^2\bar{u}_{\nu}^2\right]. \label{H}
\end{equation}
Here $u$ is a complex scalar providing via the stereographic projection a 
parametrisation of $S^2$. 
This Lagrangian provides a simple, minimal generalization of the concept of $K$ fields to a model with $S^3$ target space. Here, minimal means that only one of the degrees of freedom is of the $K$ type. For different,  non-minimal generalizations see e.g. \cite{babichev}, \cite{diaz}. Observe that the above Lagrangian can also be derived by a reduction of a scalar $K$ field coupled to the standard $SU(2)$ non-abelian gauge field i.e., of $K$-dilaton Yang--Mills theory. 
The coupling between both fields is of the 
non-minimal type and governed by the coupling function $\sigma$ which is 
chosen in the form
\begin{equation}
\sigma (\xi)=\lambda (1-\xi^2)^2. \label{sigma}
\end{equation}
Notice that the coupling function is semipositive definite and vanishes for 
two values of the scalar field $\xi_{1,2}=\pm 1$. Therefore, it plays the role 
of an effective potential with two effective vacua for $\xi$. The constant
 $\lambda$ is a free parameter of the model.
\\
The pertinent field equations read
\begin{equation}
 \partial_{\mu} \left( |\xi_{\nu} \xi^{\nu}| \xi^{\mu} \right) - 
\lambda H_{\mu \nu}^2 \xi(1-\xi^2)=0, \label{eq xi}
\end{equation}
\begin{equation}
\partial_{\mu} \left( \frac{\sigma}{(1+|u|^2)^2} K^{\mu} \right)=0 \label{eq u}
\end{equation}
where
\begin{equation}
K_{\mu}=(u_{\nu} \bar{u}^{\nu}) u_{\mu} - u_{\nu}^2 \bar{u}_{\mu}.
\end{equation}
In order to derive static solutions we introduce the coordinates
\begin{equation}
\vec{X} = \left(
\begin{array}{c}
r\sqrt{z} \cos \phi_2 \\
r\sqrt{z} \sin \phi_2 \\
r\sqrt{1-z} \cos \phi_1 \\
r\sqrt{1-z} \sin \phi_1 \\
\end{array} \right),
\end{equation}
where $z \in [0,1], \phi_1 \in [0,2\pi], \phi_2 \in [0,2\pi]$ are
coordinates on $S^3$, and $r \in \RR_+$ gives the extension to $\RR^4$. 
Moreover, we assume the ansatz
\begin{equation}
u=f(z)e^{i(n_1\phi_1 +n_2\phi_2)} \label{ansatz u}
\end{equation}
and
\begin{equation}
\xi=\xi(r). \label{ansatz xi}
\end{equation}
This ansatz provides that
\begin{equation}
\nabla \xi \nabla u = 0
\end{equation}
and, as a consequence, one can remove the coupling function $\sigma$ from 
equation (\ref{eq u}). Equation (\ref{eq u}) for the field $u$ simplifies,
in fact, to the field equation for the Lagrangian $H_{\mu\nu}^2$ with base
space $S^3$. Solutions to this model have been constructed in
\cite{ferr1}, \cite{ferr2}, and below we just review the results which we need
in the sequel.
\\
Concretely, the static equations of motion may be rewritten in the form
\begin{equation}
\frac{1}{r^3} \partial_r \left(r^3 \xi_r^3 \right)  +\nonumber
\end{equation}
\begin{equation}
+ \frac{16 \lambda f_z^2f^2}{r^4(1+f^2)^4}(n_1^2z+n_2^2(1-z))\xi (1-\xi^2)=0 
\label{eq xi1}
\end{equation}
\begin{equation}
\partial_z \left( (n_1^2z+n_2^2(1-z)) \frac{f^2f_z}{(1+f^2)^2}\right) - 
\nonumber 
\end{equation}
\begin{equation}
- (n_1^2z+n_2^2(1-z)) \frac{ff_z^2}{(1+f^2)^2}=0.
\end{equation}
The last expression may be simplified  
\begin{equation}
\partial_z \ln \left((n_1^2z+n_2^2(1-z)) \frac{ff_z}{(1+f^2)^2} \right)=0.
\end{equation}
Thus, 
\begin{equation}
\frac{ff_z}{(1+f^2)^2}=\frac{c_1}{(n_1^2z+n_2^2(1-z)) },
\end{equation}
where $c_1$ is an integration constant. One can proceed further and solve 
this equation. However, for the topologically nontrivial configurations 
the complex field $u$ should cover the whole target space $S^2$ at least once. 
This requirement gives a condition for the integration constants leading to 
the solutions 
\begin{equation}
f=\sqrt{\frac{\ln n_1^2 -\ln n_2^2}{\ln (n_1^2z+n_2^2(1-z)) -\ln n_2^2} -1}. 
\label{sol f1}
\end{equation}
In the case when $n_1= \pm n_2$ we arrive at the very simply formula
\begin{equation}
f=\sqrt{\frac{1}{z}-1}. \label{sol f2}
\end{equation}
Moreover, such a complex field being a map from $S^3$ (the base space 
parameterized by $z,\phi_1,\phi_2$ coordinates) to the target $S^2$ can be 
classified by a topological invariant known as the Hopf index. In fact, 
solution (\ref{ansatz u}), with (\ref{sol f1}), (\ref{sol f2}) is known 
to carry a non-vanishing Hopf index 
\begin{equation}
Q=n_1n_2.
\end{equation}
Let us now turn to the field equation for the real scalar $\xi$. First of 
all one can observe that this 
expression leads to an ordinary differential equation for $\xi = \xi (r)$ 
only if $n_1^2=n_2^2=n^2$ in the solution for the $u$. Therefore, only the
solution (\ref{sol f2}) is admissible. Then, the $z$-dependence in the 
second term of (\ref{eq xi1}) cancels and we get
\begin{equation}
\frac{1}{r^3} \partial_r \left(r^3 \xi_r^3 \right)  + 
\frac{4\lambda n^2}{r^4}\xi (1-\xi^2)=0. 
\end{equation}
Introducing the new variable $x=\ln r$ we find that
\begin{equation}
\xi_x^2\xi_{xx} + \frac{4\lambda n^2}{3}\xi (1-\xi^2)=0.
\end{equation}
This equation has been recently analyzed in the context of compact domain 
walls \cite{comp}. The corresponding compacton solution located at $x_0$ reads
\begin{equation} \label{sol-x}
\xi (x) = \left\{ 
\begin{array}{ll}
 -1 & \alpha x \leq \alpha x_0 -\frac{\pi}{2} \\
\sin \alpha (x-x_0) & \alpha x \in [\alpha x_0 -\frac{\pi}{2},\alpha x_0 +
\frac{\pi}{2} ] \\
1 & \alpha x \geq \alpha x_0 +\frac{\pi}{2}
\end{array}
\right.,
\end{equation}
where 
\begin{equation}
\alpha = \left( \frac{4\lambda n^2}{3} \right)^{1/4} .
\end{equation}
Finally, the 4 dimensional compacton solution is 
\begin{equation}
\xi (r) = \left\{ 
\begin{array}{ll}
 -1 & \alpha \ln r  \leq \alpha x_0 -\frac{\pi}{2} \\
\sin (\alpha \ln (r/r_0) )& \alpha \ln r \in [\alpha x_0 -\frac{\pi}{2}, \alpha x_0 
+\frac{\pi}{2} ] \\
1 & \alpha \ln r \geq \alpha x_0 +\frac{\pi}{2}
\end{array}
\right.,
\end{equation}
(where $x_0 = \ln r_0$), together with 
\begin{equation}
u (z, \phi_1,\phi_2)= \sqrt{\frac{1}{z}-1} e^{in(\phi_1+\phi_2)},
\end{equation}
with the Hopf index of the underlying Hopf maps equal to $n^2$. 
\\
The size of the compact soliton, if treated as an object living in the 
original 4 dimensional space, varies as one changes its position. The 
inner and outer compacton boundary points 
$(r_1,r_2)$ are 
\begin{equation}
 r_1=r_0e^{-\frac{\pi}{2\alpha}}, \;\;\; r_2=r_0e^{\frac{\pi}{2\alpha}},
\end{equation}
 where $x_0 = \ln r_0$ gives a parametrisation of the center of the solution. 
Thus the shell radius is
\begin{equation}
R=r_2-r_1= 2 r_0 \sinh \frac{\pi}{2\alpha}.
\end{equation}
As we see, the compacton is getting narrower as it approaches the origin. On 
the other hand its radius grows while it moves in the opposite direction.  
\\
The energy of the solution is given as follows 
\begin{equation}
E=\int dV \left( \xi_r^4 + 
(1-\xi^2)^2 \frac{16 \lambda n^2 f_z^2f^2}{r^4(1+f^2)^4}\right).
\end{equation}
Thus, 
\begin{equation}
E=\frac{(2\pi)^2}{2} \int_0^{\infty} dr \; r^3 \left( \xi_r^4 + 
\frac{4 \lambda n^2}{r^4} (1-\xi^2)^2 \right)
\end{equation}
\begin{equation}
\hspace*{0.4cm} = \frac{(2\pi)^2}{2} \int_{-\infty}^{\infty} dx \left( 
\xi_x^4 + 4 \lambda n^2(1-\xi^2)^2 \right). \label{bog}
\end{equation}
It is clearly visible that the compact solution for the real scalar field is of 
the Bogomolny type, satisfying a first order differential equation, which may 
be easily derived from (\ref{bog}) using the standard Bogomolny trick.
\\
Specifically, for the one-compacton configuration we find
\begin{equation}
E =3 \left( \frac{3}{4} \right)^{3/4} \pi^2 \lambda^{3/4} Q^{3/4}.
\end{equation} 
Interestingly, the energy depends on a non-integer power of the 
Hopf charge of the underlying 
$u$ field, like in the Vakulenko-Kapitansky formula \cite{VaKa}. 
It should be mentioned, however, that this relation between energy and Hopf
charge does not provide an energy bound for general Hopf charge, because the
full field configuration is a suspended Hopf map, and its topological
classification is therefore given by the homotopy group $\pi_4(S^3) \simeq
\ZZ_2$, as we demonstrate below.
\\
So let us prove that the obtained configuration may be understood as a 
suspended Hopf map, i.e., a map from the $S^4$ base space onto the $S^3$ target 
space characterized by the nontrivial homotopy class $\pi_4(S^3)$. It is 
convenient to combine the fields $(\xi,u)$ into a $SU(2)$ matrix $U$  
\begin{equation} \label{U}
U= \sin \frac{\pi}{2} \xi \; I + i \cos \frac{\pi}{2} \xi \; T,
\end{equation}
where
\begin{equation} \label{T}
T=\frac{1}{1+|u|^2} \left( 
\begin{array}{cc}
|u|^2-1 & -2iu \\
2i\bar{u} & 1-|u|^2
\end{array}
\right)
\end{equation}
and $I$ is the unit matrix. Thus, the $U$ field maps $\RR^4$ onto the three 
dimensional target sphere. For every fixed value of $\xi \neq \pm1$ 
the $U$ field 
is just a Hopf map $S^3 \rightarrow S^2$ with the previously found nonvanishing 
topological charge. For $\xi = \pm 1$, representing the poles of $S^3$, we 
get the identity map. Therefore we get a full covering of the $S^3$. The 
boundary condition, $U \rightarrow I$ as $r \rightarrow \infty$,  allows 
for compactification of the original $\RR^4$ space to $S^4$.  These facts 
render the $U$ a representative of the nontrivial homotopy class
\cite{williams}. We remark
that topological solitons which may be classified as suspended Hopf maps
(although not of the compacton type) have been studied recently, e.g., in
\cite{speight} and in \cite{pullback}.  
\\
Interestingly, the compactons we found do not have the structure of a ball.
Instead, they have the form of a shell, where the energy density is radially
symmetric, and is zero both inside the inner compacton boundary and outside 
the outer boundary. Further,  
the one compacton solution may be easily extended to multi-compacton 
configurations 
by taking an alternating collection of sufficiently separated compactons 
(which interpolate from the vacuum value $\xi =-1$ to $\xi =1$ with 
increasing radius)
and 
anti-compactons
(which interpolate from $\xi =1$ to $\xi =-1$ with increasing radius),
forming an onion-like structure with one compacton or
anti-compacton as the innermost shell, surrounded by further compacton and
anti-compacton shells. The energy of the solution equals just the sum of 
the energies of all $N$ compact solitons. The corresponding topological charge 
is nontrivial if the number of compactons is not the same as the number of 
anticompactons, whereas it is zero if the number of compactons and
anti-compactons is equal. We remark that the Hopf charge of the $u$ field
within each \linebreak (anti-)compacton may be chosen independently.
\\
Let us also notice that the simplest compact Hopf map is stable as far as  
linear radial perturbations are considered. In this case the  
stability analysis of \cite{comp}, \cite{comp-br1}, \cite{comp-br2} holds. 
\\
The existence of  infinitely many exact suspended 
hopfion solutions in our model makes it interesting to further investigate its integrability properties. First we observe that the model has infinitely many symmetries and, therefore, infinitely many conserved currents. The symmetries are just the area-preserving diffeomorphisms acting on the target space $S^2$ spanned by the complex scalar field $u$, and the conserved currents are the corresponding Noether currents. One way to further analyse the integrability  
is provided by  the generalized zero curvature condition of Refs.  \cite{GZC1}, 
\cite{GZC2}, which gives a well-defined extension of the standard 
integrability criterion (Zakharov-Shabat zero curvature representation) to higher 
than two dimensions. The corresponding generalized zero curvature condition 
is the condition for the holonomy in higher loop space to be independent 
of the deformations of loops or, in other words, it is just a condition for 
the flatness of the connetion in loop space. Moreover, assuming the 
reparametrization invariance of the holonomy, one gets local generalized 
zero curvature conditions, 
\begin{equation}
F_{\mu \nu}(A)=0, \;\;\; D_{\mu} B^{\mu}=0, \label{zc}
\end{equation}
i.e., flatness of a connetion $A_{\mu} \in \mathcal{G}$ and covariant
constancy of a vector field $B_{\mu} \in \mathcal{P}$, where $\mathcal{G}$ is
a Lie  algebra and $\mathcal{P}$ an abelian ideal (a representation space of 
the Lie algebra). A model is said to be integrable if one can rewrite the 
field equations as the generalized zero curvature conditions (\ref{zc}) and 
if the abelian ideal used in the construction has infinite dimensions.  
\\
One can verify that the model (\ref{model}) admits such a generalized zero 
curvature formulation provided we impose an additional constraint on the 
fields. Therefore our model, although not integrable in this sense, possesses an 
integrable sector defined by the following integrability condition
\begin{equation} \label{perp-cond}
u_{\mu} \xi^{\mu}=0.
\end{equation}
In particular, the generalized zero curvature formulation of the submodel is 
given by 
\begin{equation}
A_{\mu}=\frac{1}{1+|u|^2} \left( -iu_{\mu}
T_+-i\bar{u}_{\mu}T_- + (u\bar{u}_{\mu} -\bar{u} u_{\mu})T_3 \right)
\end{equation}
\begin{equation}
B_{\mu}=2i|\xi_{\nu} \xi^{\nu}| \xi_{\mu} \sqrt{j(j+1)} P^{(j)}_0 + 
\nonumber
\end{equation}
\begin{equation}
\hspace*{2cm} \frac{\sigma'_{\xi}}{(1+|u|^2)^3} \left(
\bar{K}_{\mu}P^{(j)}_1 + K_{\mu} P^{(j)}_{-1}
\right),
\end{equation}
where $T_{\pm},T_3$ are the generators of the $sl(2)$ Lie algebra and 
$P^{(j)}_m$ transforms under the spin-$j$ representation of $sl(2)$. The 
equations of motion for the submodel are given by GZC in any spin 
representation, which implies the generalized integrability. The importance 
of this submodel emerges from the fact that our hopfions belong to it. 
Indeed, they solve the field equation together with the constraint. 
\\
One can compare the integrability properties of our model with the most known example of $S^3$ target space model i.e., the Skyrme model. The Skyrme model does not have infinitely many symmetries, but it also has an integrable sector. However, in this case it is given by two integrability conditions
\begin{equation}
 u_{\mu} \xi^{\mu}=0, \;\;\; u_{\mu}^2=0.
\end{equation}
Thus, the integrable submodel is much more constrained, and this fact obviously affects the chance for the existence of exact solitons. In particular, the new condition, that is, the eikonal equation, is rather restrictive. It leads to one exact, given solution once the nontrivial topology (the Hopf charge) for $u$ is fixed (more precisely, it leads to one fixed solution up to Moebius transformations, that is, rotations of the target space $S^2$ where $u$ takes its values). This is in contrast  to the first integrability condition, which is easily obeyed by a general separation of variables ansatz. This can explain why our model as well as its extention presented in the next section possess infinitely many exact solutions based on infinitely many Hopf maps
(or infinitely many maps $S^2 \to S^2$ in the next section).  
\\
The relation between the restrictions imposed by the integrability conditions and existence of exact solutions is also confirmed by recent investigations of the pure quartic (Euclidean) Skyrme model by Speight [22]. In fact, in this model, which from the point of view of the generalized integrability is exactly the same as the Skyrme model, only one exact soliton with topology of the suspended hopfion has been found. It is not of the compacton type.

\section{Some 3-dimensional compactons} 

Here we want to study briefly some models which give rise to compacton
solutions analogous to the case discussed in the preceding section, but
in 3+1 dimensions, which is the case more directly relevant for physical
applications. We shall discuss explicitly two cases.

\subsection{Shell-shaped compactons}
In the first example, we observe that the Lagrangian (\ref{model}) of Section
2 is quartic in first derivatives, therefore it is scale invariant precisely
in four dimensions, which is one way to circumvent Derrick's theorem and have
static solutions in four dimensions. If we want to have static solutions in
three dimensions, one possibility consists, therefore, in choosing a
Lagrangian cubic in first derivatives. This implies, however, that the
resulting Lagrangian is non-polynomial. 
For models of this type the study of time-dependent dynamics is
problematic (e.g. boundedness of the energy, or global hyperbolicity),
therefore we shall introduce the energy functional for static configurations
directly. 
Concretely, the  three-dimensional model we study 
has the following
energy functional for static configurations
\begin{equation}
 E =\int d^3 x [\left(  \xi_k \xi_k\right)^\frac{3}{2} -  \left( \sigma (\xi)
H_{jk}^2\right)^\frac{3}{4} ]
\label{model1-3d}
\end{equation} 
where $j,k = 1,2,3$. It is related to the model (\ref{model}) of Section 2
such that both terms of the model (\ref{model}) are taken to the power 
$\frac{3}{4}$. Further, we have already reduced to the static case.
If we now introduce three-dimensional spherical polar coordinates
$(x_1 ,x_2 ,x_3) \to (r,\theta ,\varphi )$ and use the ansatz
$\xi = \xi (r)$, $u = u(\theta ,\varphi )$, then the coupling function can
again be removed from the equation for $u$, and this equation can be written as
\begin{equation}
\partial_j \left( \frac{K_j [(u_l\bar u_l)^2 - u_l^2 \bar
    u_k^2]^{-\frac{1}{4}}}{(1+u\bar u)} \right) =0 .
\end{equation}
This equation is just the field equation of the model of Aratyn, Ferreira and
Zimermann (AFZ)\footnote{The energy density $(H_{jk}^2)^\frac{3}{4}$ is, 
in fact, precisely the energy density of the AFZ
model. In three-dimensional, Euclidean base space, the AFZ model has infinitely
many soliton solutions of the knot type \cite{AFZ1}, \cite{AFZ2}, whose
existence is related both to the conformal base space symmetry and to
the infinitely many target space symmetries of this model. 
Here we are, however, interested in
solutions on the base space $S^2$.}.
For the ansatz $u(\theta ,\varphi)$ it has the solutions
\begin{equation}
u = \tan \frac{\theta}{2} e^{in\varphi}
\end{equation}
where $n$ is an integer and these solutions $u$ describe maps $S^2 \to S^2$
with winding number $n$. The corresponding $H_{jk}^2$ reads
\begin{equation}
H_{jk}^2 = \frac{n^2}{4 r^4} .
\end{equation}
The equation for $\xi (r)$ for this ansatz is
\begin{equation}
3  \frac{1}{r^2} \partial_r (r^2 \xi_r^2 ) + \frac{3}{4}
(H_{jk}^2)^\frac{3}{4} \sigma^{-\frac{1}{4}} \sigma_\xi =0
\end{equation}
or, for the specific coupling function $\sigma = \lambda (1-\xi^2)^2$ and the 
$H_{jl}^2$ above,
\begin{equation}
 \frac{1}{r^2} \partial_r (r^2 \xi_r^2 ) + 
\left( \frac{\lambda n^2}{4}\right)^\frac{3}{4} \frac{1}{r^3} \xi 
(1-\xi^2)^\frac{1}{2} =0.
\end{equation}
Introducing again the variable $x=\ln r$, this equation becomes
\begin{equation}
\xi_x \xi_{xx} + 2^{-2}\left( \lambda n^2
\right)^\frac{3}{4} \xi (1-\xi^2)^\frac{1}{2} =0
\end{equation}
which has exactly the same compacton solution (\ref{sol-x}) as in Section 2,
where now the constant $\alpha$ is
\begin{equation}
\alpha = 2^{-\frac{2}{3}}\left( \lambda n^2
\right)^\frac{1}{4}
\end{equation}
Therefore, this model has exactly the same shell-like 
spherically symmetric compacton solutions in
three dimensions as the previous model of section 2 has in four dimensions.
\\
The topology of the compacton solutions is now given by maps $S^3 \to S^3$.
This topology may again be described by the SU(2) matrix $U$ of Eq. 
(\ref{U}), where, however, now the complex scalar $u$ is a map $S^2 \to S^2$.
Therefore, the solution field configuration takes its value at the north pole
of the target space $S^3$ for $\xi =-1$ (inside the inner shell boundary),
covers the full target $S^3$ while $\xi$ varies from $\xi = -1$ to $\xi =1$, 
and takes its value at the south pole for $\xi =1$ (outside the outer shell
boundary).  
\\
Also the integrability properties are exactly equivalent to the ones of the model of Section 2.
The present model has infinitely many symmetries and infinitely many conserved currents,
like the one of Section 2. Further, the generalized zero curvature representation does not exist 
for the full model, but only for a submodel defined by the additional condition (\ref{perp-cond}),
again in complete analogy with the model of Section 2.   

\subsection{Ball-shaped compactons}
Another simple modification which allows for compactons 
in three spatial dimensions is given by the
Lagrangian
\begin{equation}
L= |\xi_{\nu} \xi^{\nu}| \xi_{\mu} \xi^{\mu} - \sigma (\xi)  \bar H. 
\label{model2-3d}
\end{equation}
Here, the term 
\begin{equation}
\bar H \equiv \frac{u_\mu \bar u^\mu}{(1+u\bar u)^2}
\end{equation}
is just the Lagrangian of the CP$^1$ model. The above Lagrangian contains one
quartic term and one quadratic term in first derivatives, and so may have
finite energy solutions in three dimensions, according to the Derrick scaling
argument. 
It is, however, not scale invariant nor does it have infinitely many
symmetries, in contrast to the models studied above. Therefore, we do not
expect to find fully analytical solutions in this case, see below.
\\
We again use the ansatz $\xi =
\xi (r)$, $u = u(\theta ,\varphi)$ in spherical polar coordinates in three
space dimensions. With this ansatz, the coupling function $\sigma = \lambda
(1-\xi^2)^2$ may again be eliminated from the field equation for $u$, and this
equation is, therefore, just the field equation of the CP$^1$ model on base
space $S^2$. The simplest solution of this equation is
\begin{equation} \label{simp-sol}
 u = \tan \frac{\theta}{2} e^{i\varphi} .
\end{equation}
The CP$^1$ energy density of this solution is
\begin{equation}
-\bar H = \frac{\nabla u \cdot \nabla \bar u}{(1+u\bar u)^2} = \frac{1}{2r^2}.
\end{equation}
There exist many more solutions of the CP$^1$ model like, e.g., higher powers
of the simplest solution, but the corresponding energy densities are no longer
independent of the angular coordinates. These higher solutions are, therefore,
not compatible with our separation ansatz $\xi = \xi(r)$, and we
have to restrict to the simplest solution (\ref{simp-sol}) in what follows.  
For this simplest
CP$^1$ solution, we find the following Euler--Lagrange equation for $\xi (r)$
\begin{equation}
\frac{1}{r^2} \partial_r \left( r^2 \xi_r^3 \right) + \frac{\lambda}{2r^2}
\xi (1-\xi^2 )=0
\end{equation}
or, after the variable transformation $s=r^\frac{1}{3}$
\begin{equation} \label{s-eq}
\frac{1}{27} \xi_{ss} \xi_s^2 + \frac{\lambda}{2} s^2 \xi (1-\xi^2) =0 .
\end{equation}
This equation differs from the previous ones by the explicit presence of the 
factor $s^2$ (the independent variable) in the second term, that is, it is no
longer an autonomous equation. As expected, we were not able to find 
analytic solutions to this equation, so we will resort to a qualitative
analysis and to a numerical study in the sequel. Further,
we shall find that its solutions are
no longer topological and may, therefore, have arbitrarily small energies.
\\
Concretely, a numerical integration of Eq. (\ref{s-eq}) leads to the following
results: 
\begin{itemize}
\item 
There do not exist shell-type solutions. If one starts the 
integration 
at an inner boundary $s_0 >0$ with $\xi (s_0)$ taking one vacuum value (e.g.
$\xi (s_0)=-1$), and $\xi' (s_0)=0$, then the integration into the direction
$s>s_0$ never reaches the other vacuum value $\xi =+1$. Instead, a point $s_1
> s_0$ is reached where $\xi'(s_1)=0$ and $-1 < \xi (s_1) < 1$, and at this
point $\xi (s)$ becomes singular (it is obvious from Eq. (\ref{s-eq}) that
at a point where $\xi ' =0$, either $\xi$ must take one of its vacuum values,
or $\xi''$ becomes singular).
\item
There exist, however, solutions of the ball type. If one starts the
  numerical integration at an outer compacton boundary (e.g., with 
$\xi (s_1)=+1$ and $\xi' (s_1)=0$) and integrates towards $s<s_1$, then the
integration will simply hit the point $s=0$. In order to see that the
resulting solution is an acceptable compacton, it is more useful to reverse
the integration and to start at $s=0$.
\item
 Let us assume that we start the integration at $s=0$ with some value
$0 < \xi (0) < 1$ and with $\xi '(0) =k >0$. First, we observe that due to the
suppression factor $s^2$ in the second term of Eq. (\ref{s-eq}), $\xi (0)$ and
$\xi' (0)$ may take arbitrary values without making $\xi'' (0)$ singular
(concretely, if $\xi '(0) >0$ then $\xi ''(0)=0$). 
For $s > 0$, 
we note that for $0 < \xi < 1$ it holds that $\xi '' <0$, whereas for
$\xi >1$ it holds that $\xi ''>0$, as follows easily from Eq. (\ref{s-eq}).
\item
Therefore, with the initial conditions given above,
the following picture emerges for an integration starting at
$s=0$. If $k \equiv \xi' (0) >0$ is too large, then the integration curve for
$\xi (s)$ will cross the line $\xi =1$ and then grow forever, producing a
formal solution with infinite energy. If $k$ is too small, the integration
curve will reach a point $s_2$ where $\xi'(s_2)=0$ but still $\xi (s_2)<1$. 
At this point the integration curve becomes singular, because $\xi'' (s_2)$ is
singular. It follows that there exists a fine tuned value $k_*$ for the
integration constant $k>0$ such that
the integration curve touches the line $\xi =1$ instead of crossing it, that
is, it reaches the value  $\xi'(s_1)=0$ precisely at the point $s_1$ where
$\xi (s_1) =1$. This configuration is the compacton.
The above qualitative discussion is completely confirmed by an
explicit numerical integration.
\end{itemize}
In the above argument, we could start the integration at $s=0$ for an
arbitrary value $0 < \xi (0) <1$. By choosing a $\xi (0)$ arbitrarily close
to the value $+1$ we can, therefore, make the size and the energy of the
compacton arbitrarily small. These compactons are, therefore, no longer
topological. This makes their stability under time-dependent perturbations
more problematic (a detailed stability analysis is beyond the scope of the
present letter).
\\
{\em Remark:} We chose the coupling function $\sigma =\lambda (1-\xi^2)^2$ as
the simplest representative of a class of coupling functions with (at least)
two vacuum values in order to allow for topological compactons. In the last
example, however, the compactons are not topological in any case, therefore
the presence of more than one vacuum in the coupling function is not necessary
in this case.
\\
{\em Remark:} In the above 
discussion about the integration from the center $s=0$
we restricted to the interval $0 < \xi (0) < 1$ just for reasons of
simplicity. It presents no difficulty to extend the discussion and to cover
cases where $\xi$ starts outside this interval at $s=0$. For an adequately
fine-tuned value of $k=\xi '(0)$ there always exists a compacton. We plot two typical compact ball
solutions in Figures 1 and 2. Here in Figure 1 $\xi$ starts at $\xi (0)=0.465$ whereas in Figure 2 it starts at $\xi (0) = -0.833$. The qualitative discussion above is completely confirmed by the numerical solutions shown in Figures 1 and 2.
  
\section{Discussion}
In this paper, we found analytic solutions for compact solitons of the shell type in 3+1 and  4+1-dimensional $K$ field theories.
To our knowledge, these are the first examples of compactons in $K$ field theories in those dimensions. The topology of these shell type compactons is quite interesting, being a combination of the topology of the complex field $u$ which lives on a compact base space ($S^2$ or $S^3$, respectively) and is classified by the homotopy of maps from that base space to the target space $S^2$, and of the topology of the real field $\xi$ which lives on the positive real line (radial coordinate), and where the topology is induced by the boundary conditions implied by the finiteness of the energy. This situation is reminiscent of the Skyrme model, which has exactly the same field contents and the same topology. The main difference is the quartic kinetic term of the real scalar field $\xi$ in the models discussed in the present paper. Further, we were able to find an infinite number of exact compacton solutions in both cases, labelled by an integer $n$. The existence of these infinitely many analytical solutions is probably related to the infinitely many symmetries and infinitely many conserved currents, which are present in both models, as typically happens in models with infinitely many conservation laws. 
We also investigated the integrability properties of these two models and found that, although they have infinitely many conservation laws, they do not possess a generalized zero curvature representation, but a submodel defined by a further constraint equation does. In this sense, these models seem to be half way between integrable and non-integrable. 
\\
Finally we investigated a model without infinitely many symmetries, and without the scale symmetry of the models discussed above. This model does not allow for analytical solutions, but it is not difficult to find solutions numerically. It turns out that this model does not have solutions of the shell type. It does, however, possess compact solutions of the ball type, but these solutions are no longer topological solitons. The real scalar field $\xi$ may, in fact, take values on arbitrarily small intervals $[\xi_0,1]$ of the real axis (here $\xi_0$ is the starting value at zero radius, and 1 is the vacuum value), so the full target space may be an arbitrary segment of $S^3$. Consequently, there exist compact balls with arbitrary size and energy.  
\\
As stated, it has been the main purpose of the present paper to establish the existence of compactons in $K$ field theories in higher dimensions, and to study their properties.
In the case of the shells the underlying theories are scale invariant, and
the dynamical generation of a scale (the compacton size) is by itself an important result in relativistic field theory.
Nevertheless, we want to discuss briefly some possible applications and further directions of investigation. As already mentioned, compactons in low (i.e. 1+1) dimensions have been applied to brane cosmology, where the compacton acts as a domain wall such that the confinement of both gravity and field fluctuations to the domain wall ("thick brane") is an automatic consequence of the field equations. Recently, compact solutions of the $Q$-ball type and of the $Q$-shell type have been found in certain higher (i.e. 3+1 dimensional)
gauge theories with normal kinetic terms and V-shaped potential (for the
scalar field), \cite{ArLi}. Further, in Ref. \cite{KKLL} the coupling of these
compact $Q$ balls and $Q$ shells to gravity has been investigated in detail by
the use of numerical methods. It was found there that the $Q$ balls may form
stable self-gravitating solutions, whereas the $Q$ shells may not only form
self-gravitating $Q$ shells but may also dress black holes (i.e. may contain a
Schwarzschild interior). This whole analysis is complicated by the fact that
the basic solutions of Ref. \cite{ArLi} are (time dependent) $Q$ balls and $Q$
shells rather  than static solitons (e.g. the asymptotic metric is Kerr rather
than Schwarzschild). Therefore, the coupling of the solutions of the present paper (or of related solutions of comparable simplicity) to gravity should lead to a simpler analysis of the problem of gravitational self-coupling and probably even allow for  mainly analytical calculations. This question is under current investigation. 

\section*{Acknowledgements}
C.A., P.K. and J.S.-G. thank MCyT (Spain) and FEDER
(FPA2005-01963), and support from
 Xunta de Galicia (grant PGIDIT06PXIB296182PR and Conselleria de
Educacion). A.W. acknowledges support from the Foundation for
Polish Science FNP (KOLUMB programme 2008/2009) and Ministry of
Science and Higher Education of Poland grant N N202 126735
(2008-2010).

\vspace*{15cm}

\newpage

\begin{figure}[h!]
\begin{center}
\includegraphics[width=0.65\textwidth]{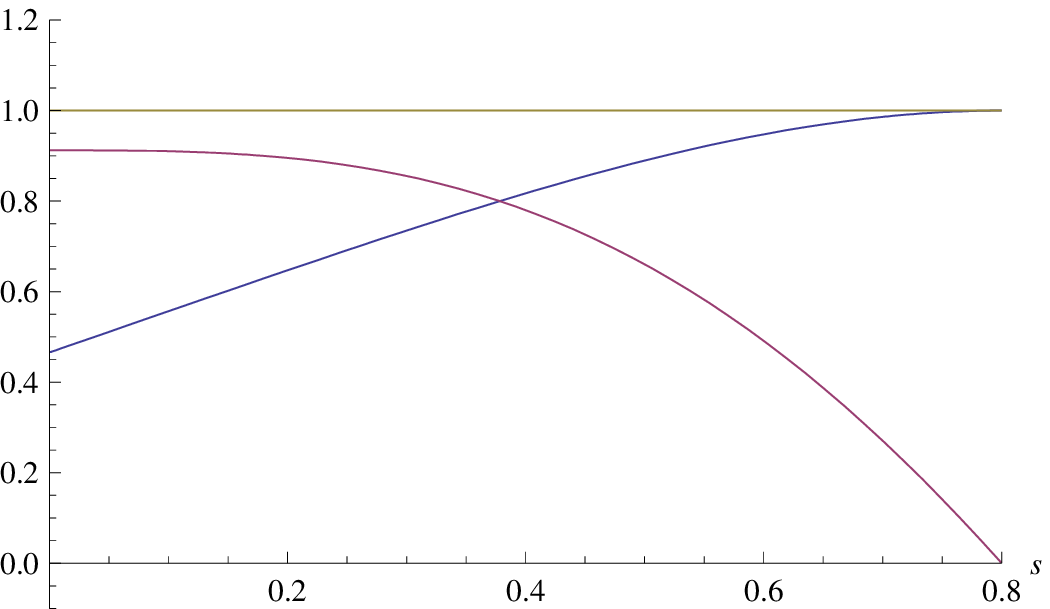}
\caption{The compact ball of Section 3.2. The starting values for the field $\xi$ are $\xi(0)=0.465$, $\xi'(0)=0.912$, and the resulting compacton radius is $s_1 = 0.8$. We display both the compacton field $\xi$, starting at $0.465$ and ending at the vacuum value 1, and the first derivative $\xi'$.}
\end{center} 
\end{figure}

\begin{figure}[h!]
\begin{center}
\includegraphics[width=0.65\textwidth]{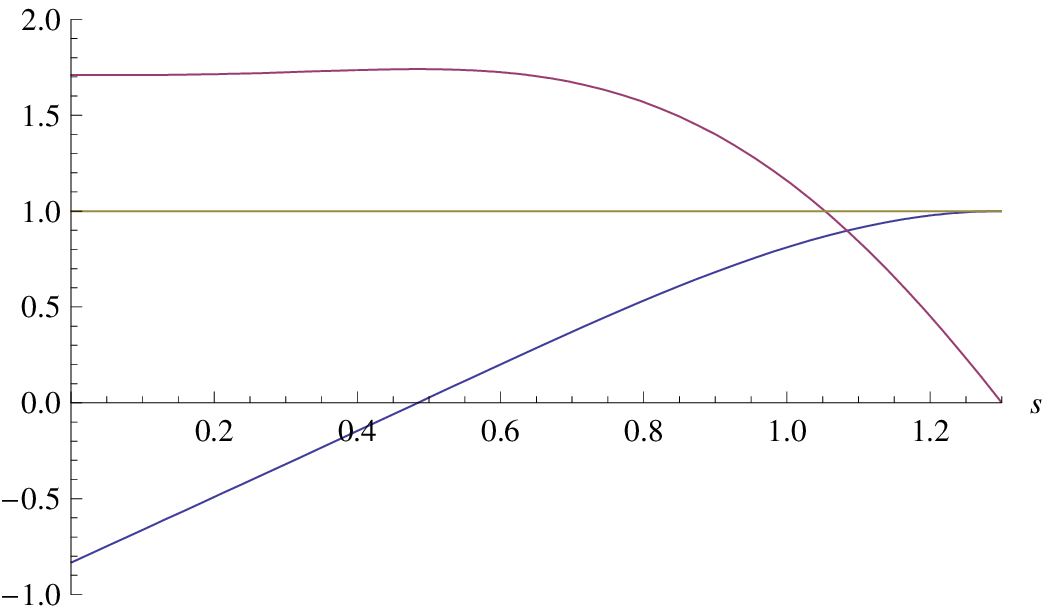}
\caption{The compact ball of Section 3.2. The starting values for the field $\xi$ are $\xi(0)=-0.833$, $\xi'(0)=1.709$, and the resulting compacton radius is $s_1 = 1.3$. We display both the compacton field $\xi$, starting at $-0.833$ and ending at the vacuum value 1, and the first derivative $\xi'$.}
\end{center} 
\end{figure}

\end{document}